\documentclass[twocolumn,showpacs,preprintnumbers,amsmath,amssymb]{revtex4}
\usepackage{amssymb}
\usepackage[dvips]{graphicx}

\begin{document}

\author{T. Mertelj$^{1,2}$, V.V. Kabanov$^{1,3}$, J. Miranda$^{1}$  and D. Mihailovic$^{1,2}$}
\date{\today}

\affiliation{$^{1}$Jozef Stefan Institute, Jamova 39, 1000
Ljubljana, Slovenia} \affiliation{$^{2}$Faculty of Mathematics and
Physics, University of Ljubljana, Slovenia} \affiliation{$^{3}$
Department of Physics, Loughborough University, Loughborough LE11
3TU, UK}

\begin{abstract}
Self-organization of charged particles on a 2D lattice, subject to
an anisotropic Jahn-Teller-type interaction and 3D Coulomb repulsion is
investigated. In the mean-field approximation without Coulomb interaction,
the system displays a phase transition of first order. In the presence of
the Coulomb repulsion
the global phase separation becomes unfavorable and the system shows
a mesoscopic phase separation, where the size of the charged regions
is determined by the competition between the ordering energy and
the Coulomb energy.

The phase diagram of the system as a function of particle density
and temperature is obtained by systematic Monte Carlo simulations.
With decreasing temperature a crossover from a disordered state to
a state composed from mesoscopic charged clusters is observed. In
the phase separated state charged clusters with even number of
particles are more stable than those with odd number of particles
in a large range of particle densities. With increasing particle
density at low temperatures a series of crossovers between states
with different cluster sizes is observed. Above half filling in
addition to the low temperature clustering another higher
temperature scale, which corresponds to orbital ordering of
particles, appears.

We suggest that the diverse functional behaviour - including
superconductivity - observed in transition metal oxides can be
thought to arise from the self-organization of this type.
\end{abstract}

\title{Self-organization of charged particles on a 2D lattice subject to
anisotropic Jahn-Teller-type interaction and 3D Coulomb repulsion}
\maketitle

\section{Introduction}

The presence of nanoscale inhomogeneities is ubiquitous in the
cuprate superconductors\cite
{EgamiPintschovius,BillingeBianconi,Davis,Demsar,screview}, the
magnetoresistive manganites\cite
{DeteresaIbarra97,AllodiRenzi97,UeharaMori99,FathFreisam99,
PapavassiliouFardis2000,KwasakiMinami2006} and other doped
transition metal oxides\cite{KuhnsHoch03, SageBlake2006,
CheongHwang1994}. Furthermore, there is emerging consensus that
doped charge carriers in the oxides may phase segregate to form
nano-scale textures. These are believed to be of importance for
achieving their functional properties such is superconductivity in
the cuprates\cite{screview} and giant magnetoresistance in the
manganites\cite {Dagotto2001}.

For the cuprates the idea of charge segregation appeared soon
after the discovery of superconductivity
\cite{zaanen,emery,gorkov}. In a doped semiconductor the phase
separation may have two different origins. The first is the
chemical origin and is associated with the segregation of dopant
atoms. This type of phase segregation is usually temperature
independent and weakly dependent on external perturbation.
Exceptions may appear due to a large mobility of dopant atoms at
relatively high temperature.

If the mobility of impurity atoms is small one might expect a
pure electronic mechanism of phase separation. In this case the
electronic system is in thermodynamic equilibrium and competing
phases are close in energy. This is typical for the systems
exhibiting a first order phase transition. Electronic phase
separation is very often observed in the magnetic semiconductors
like $EuSe$ or $EuTe$ \cite{vitins,shapira1, shapira2}. Therefore
the idea of the charge segregation in the cuprate superconductors
and in the manganites is very often associated with magnetic
degrees of freedom \cite{emery90,emery93,uhrig}, where the phase
separation is discussed within $t-J$ model. In Refs.
\cite{angelucci,singh,bang,moreo} phase separation was studied
within Hubbard model. The results are still controversial. In some
cases the $t-J$ model displays clear static \cite{emery90} or
dynamic\cite{emery93} phase separation. The situation is quite
different for the Hubbard model. For example the results of
numerical simulations\cite{moreo} suggest that the phase
separation is absent at any set of parameters and for any size of
the lattice. Nevertheless, all these models do not consider {\it
long-range} Coulomb repulsion which has very strong effect on the
phase separation.
\cite{gorkov,Low,lorenzana,MerteljKabanov2005,spivak,ortix,muratov}

The long-range Coulomb repulsion together with the surface energy
determine the topology of the two phase state. The charged
carriers have the tendency towards spatial segregation which is
caused by the fact that the free energy density of the phase with
finite density of carriers is lower then the free energy density
of the undoped system. On the other hand, the charge segregation leads to
the charging effect because the dopant atoms are distributed
uniformly in the system. Therefore, a strong electric field
appears which has tendency to mix the charged phases. In the
low doping limit there is a low concentration of charged droplets and
they do not overlap. The system behaves as an insulator. When the
concentration increases the percolative transition to a new phase
is expected\cite{mkm,Dagotto2001,abk}.

More recently it was suggested that an interplay of a short range
lattice attraction and the long-range Coulomb repulsion could lead
to the formation of short metallic or insulating strings of
polarons\cite{fedia1,akpz}. This was mainly motivated by
observation of giant isotope effect in manganites and
cuprates\cite{guomeng1,guomeng2}. In ref. \cite{mk} we suggested
that an anisotropic mesoscopic Jahn-Teller interaction between
electrons and $k\neq 0~$ optical phonons might lead to the
formation of carrier pairs and stripes. A slightly different
approach, based on elasticity was considered more recently for the
case of the manganites by Kugel and Khomskii \cite{klim} using the
methods of Eremin et al.\cite{eremin}, and by Shenoy et
al.\cite{Shenoy}.

The fundamental question which we try and answer here is how
charged particles order in the presence of anisotropic Jahn-Teller
type interaction, particularly when their density becomes large.
We consider charged particles on a 2D square lattice subject to
\textit{only} the long-range Coulomb interaction and an
anisotropic Jahn-Teller (JT) deformation. In the preliminary
report we have considered a narrow doping range, but have found a
clear evidence of phase segregation and preferential formation of
pairs.\cite{MerteljKabanov2005} Here we extend this study over the
full doping range.

In the mean field (MF) approximation without Coulomb repulsion,
the system displays a first order phase transition to an ordered
state below some critical temperature. In the presence of Coulomb
repulsion global phase separation becomes unfavorable and the
system shows a mesoscopic phase separation, where the size of
charged regions is determined by the competition between the
ordering energy and the Coulomb energy. Using Monte-Carlo (MC)
simulations we show that the system can form many different
mesoscopic textures, such as clusters and stripes, depending only
on the magnitude of the Coulomb repulsion compared to the
anisotropic lattice attraction and the density of charged
particles. Surprisingly, in agreement with previous report a
feature arising from the anisotropy introduced by the Jahn-Teller
interaction is that in a wide part of the phase diagram objects
with even number of particles are found to be more stable than
with odd number particles, which could be significant for
superconductivity when tunnelling is included\cite{mkm}.

\section{Formulation}

The model proposed in the ref.\cite{mk} involves all interactions
allowed by the symmetry. We consider a simplified version of the
model, where only the interaction leading to the deformation of
the $B_{1g}$ symmetry is taken into account. The interaction with
$B_{2g}$ mode leads to similar effects and therefore for our
purposes we can restrict ourselves by consideration $B_{1g}$ mode
only. As a result the interacting part of the Hamiltonian has the
form:
\begin{equation}
H_{JT}=g\sum_{\mathbf{r},\mathbf{l}}\sigma _{3,\mathbf{l}%
}\{(r_{x}^{2}-r_{y}^{2})f_{0}(r)\}(b_{\mathbf{l+r}}^ {\dagger
}+b_{\mathbf{l+r}}),
\end{equation}
here the Pauli matrix $\sigma_{3,\mathbf{l}}$ describes two
components of the electronic doublet, and $f_{0}(r)$ is a
symmetric function describing the range of the interaction. We
omit the spin index in the sum, since we ignore spin structure at
present. The resulting model could be easily reduced to a lattice
gas model. This is performed using the Lang-Firsov transformation
or equivalently the adiabatic approximation for the phonon field.
Let us introduce the classical variable $\Phi _{\mathbf{i}
}=(b_{\mathbf{i}}^{+}+b_{\mathbf{i}})/\sqrt{2}$ and minimize the
energy as a function of $\Phi _{\mathbf{i}}$ in presence of the
harmonic term $\omega \sum_{\mathbf{i}}\Phi _{\mathbf{i}}^{2}/2$.
We obtain the deformation, which corresponds to the minimum
energy,
\begin{equation}
\Phi _{\mathbf{i}}^{(0)}=-\sqrt{2}g/\omega \sum_{\mathbf{r}}\sigma
_{3, \mathbf{i}+\mathbf{r}}f(\mathbf{r}),
\end{equation}
where $f(\mathbf{r})=(r_{x}^{2}-r_{y}^{2})f_{0}(r)$. Substituting
$\Phi _{ \mathbf{i}}^{(0)}$ to the Hamiltonian (1) and taking into
account that the carriers are charged we arrive to the lattice gas
model. To formulate the model we use a pseudospin operator $S$
with $S=1$ to describe the occupancies of the two electronic
levels $n_{1}$ and $n_{2}$. Here $S^{z}=1$ corresponds to the
state with $n_{1}=1$ , $n_{2}=0$, $S_{i}^{z}=-1$ to $ n_{1}=0$,
$n_{2}=1$ and $S_{i}^{z}=0$ to $n_{1}=n_{2}=0$. Simultaneous
occupancy of both levels is excluded due to the high onsite
Coulomb repulsion (CR) energy. The Hamiltonian in terms of the
pseudospin operator is given by
\begin{equation}
H_{JT-C}^{LG}=\sum_{\mathbf{i},\mathbf{j}}(V_{l}(\mathbf{i}-\mathbf{j})S_{
\mathbf{i}}^{z}S_{\mathbf{j}}^{z}+V_{c}(\mathbf{i}-\mathbf{j})Q_{\mathbf{i}
}Q_{\mathbf{j}}),  \label{eq_HJT}
\end{equation}
where $Q_{\mathbf{i}}=(S_{\mathbf{i}}^{z})^{2}$. $V_{c}(\mathbf{m}
)=e^{2}/\epsilon _{0}am$ is the Coulomb potential, $e$ is the
charge of electron, $\epsilon _{0}$ is the static dielectric
constant and $a$ is the effective unit cell period. The
anisotropic short range attraction potential is given by
\begin{equation}
V_{l}(\mathbf{m})=g^{2}/\omega
\sum_{\mathbf{i}}f(\mathbf{i})f(\mathbf{m}+ \mathbf{i}).
\end{equation}

The attraction in this model is generated by the interaction of
electrons with optical phonons. The radius of the attraction force
is determined by the radius of the electron-phonon interaction and
the dispersion of the optical phonons\cite{akpz}.

A similar model can be formulated in the limit of the continuous
media. In this case the deformation is characterized by components
of the strain tensor. For the two dimensional case we can define 3
components of the strain tensor: $e_{1}=u_{xx}+u_{yy}$
transforming as the $A_{1g}$ representation of the $D_{4h}$ group,
$\epsilon =u_{xx}-u_{yy}$ transforming as the $B_{1g}$
representation and $e_{3}=u_{xy}$ transforming as the $B_{2g} $
representation. These components of the tensor are coupled
linearly with the two-fold degenerate electronic state which
transforms as the $E_{g}$ or $E_{u}$ representation of the point
group. Similarly to the case of interaction with optical phonons
we will keep the interaction with deformation of the $B_{1g}$
symmetry, namely $\epsilon $ only. The Hamiltonian without the
Coulomb term has the form:
\begin{equation}
H=\tilde{g}\sum_{\mathbf{i}}S_{\mathbf{i}}^{z}\epsilon
_{\mathbf{i}}+\frac{1}{2} \left(
A_{1}e_{1,\mathbf{i}}^{2}+A_{2}\epsilon
_{\mathbf{i}}^{2}+A_{3}e_{3, \mathbf{i}}^{2}\right)
\end{equation}
here $A_{j}$ are corresponding components of the elastic modulus
tensor, and $\tilde{g}$ is coupling constant of the charge
carriers with the strain tensor. The components of the strain
tensor are not independent \cite{Shenoy} and obey the
compatibility condition:
\[
\nabla^{2} e_{1}({\bf r})-4\partial^{2} e_{3}({\bf r})/\partial x
\partial y = (\partial^{2}/\partial x^{2} - \partial^{2}/\partial
y^{2}) \epsilon({\bf r}).
\]
 The compatibility condition leads to the
long range anisotropic interaction between polarons. To derive the
Hamiltonian we minimize Eq.(5) with respect to $e_{1}$ and $e_{3}$
taking into account compatibility condition. The resulting
Hamiltonian in the reciprocal space has the form:
\begin{equation}
H=\tilde{g}\sum_{\mathbf{k}}S_{\mathbf{k}}^{z}\epsilon
_{\mathbf{k}}+(A_{2}+A_{1}U( \mathbf{k}))\frac{\epsilon
_{\mathbf{k}}^{2}}{2}.
\end{equation}
The wavevector dependence of the potential is given by
\begin{equation}
U(\mathbf{k})=\frac{(k_{x}^{2}-k_{y}^{2})^{2}}{%
k^{4}+8(A_{1}/A_{3})k_{x}^{2}k_{y}^{2}}\text{.}
\end{equation}
By minimizing the energy with respect to $\epsilon _{k}$ and including
 the long-range CR we again obtain Eq.(3). The anisotropic
interaction potential $V_{l}(\mathbf{m})=-\sum_{\mathbf{k}}\exp
{(i\mathbf{ k\cdot
m})}\frac{\tilde{g}^{2}}{2(A_{2}+A_{1}U(\mathbf{k}))}$ is
determined in this case by the interaction with the classical
deformation and is long-range as well. It decays as $1/r^{2}$ at
large distances. Since attraction forces decay faster then the
Coulomb repulsion at large distances the attraction can overcome
the Coulomb repulsion at short distances leading to the mesoscopic
phase separation.

Irrespective of whether the resulting interaction between polarons
is generated by acoustic or optical phonons the main physical
picture remains the same. In both cases there is an anisotropic
attraction between polarons over short distances. This interaction
can be either ferromagnetic or antiferromagnetic in terms of the
pseudospin operators depending on the spatial direction. Without
loosing generality we assume that $V(\mathbf{m})$ is nonzero only
for the nearest neighbors and can be either ferromagnetic or
antiferromagnetic.

\section{Mean Field}

Our main goal is to study this lattice gas model (3) at a constant
average density,
\begin{equation}
n=\frac{1}{N}\sum_{\mathbf{i}}Q_{\mathbf{i}},  \label{eq_npart}
\end{equation}
where $N$ is the total number of sites. However, to clarify the
physical picture we first perform calculations in absence of
long-range CR at a fixed chemical potential first by adding the
term $-\mu \sum_{\mathbf{i}}Q_{\mathbf{i}}$ to the Hamiltonian
(\ref{eq_HJT}).

Similar models were studied many years ago on the basis of the
molecular-field approximation in the Bragg-Williams formalism
\cite{lajz,siva}. The mean-field equations for the two variables
$n$ and $M= \frac{1}{N}\sum_{\mathbf{i}}S_{\mathbf{i}}^{z}$ have
the form\cite{lajz}:
\begin{equation}
M=\frac{2\sinh {(2zV_{l}M/{k}_{B}{T})}}{\exp {(-\mu
/k}_{B}{T)}+2\cosh { (2zV_{l}M/{k}_{B}{T})}},  \label{eq_M}
\end{equation}
\begin{equation}
n=\frac{2\cosh {(2zV_{l}M/{k}_{B}{T})}}{\exp {(-\mu
/{k}_{B}T)}+2\cosh { (2zV_{l}M/{k}_{B}{T})}}  \label{eq_nvsM}
\end{equation}
here $z=4$ is the number of the nearest neighbours for the square
lattice in 2D and ${{k}_{B}}${\ is the Boltzman constant}. For
positive $\mu >0$ equation (\ref{eq_M}) has 2 solutions bellow
$T_{c}$. The solution with $M=0$ is unstable while the solution
with a finite $M$ corresponds to the global minimum with $n\to 1$
for $T\to 0$. When $-2zV_{l}<\mu <0$ the equation has 3 solutions
below $T_{c1}<T_{c}$. The free energy has two minima and one
maximum. The phase transition at $T_{c1}$ is of first order. The
trivial solution $M=0$ corresponds to the case when $n\to 0$ as
$T\to 0$. For $\mu <-2zV_{l}$ there is only the trivial solution
of the equation $M=0$.

When the number of particles is fixed (Eq.(\ref{eq_npart})) the
system is unstable with respect to global phase separation below
$T_{crit}(n)$. The line of the phase transition is determined by
the condition: $F(M=0,\mu _{crit}(T),T)=F(M(T),\mu _{crit}(T),T)$
where $F\,$is the free energy, $\mu _{crit}(T)\,\,$is the critical
chemical potential and $M$ is the solution of Eq. (\ref{eq_M}). As
a result, at a fixed average $n$ two phases with
$n_{0}(T)=n(M=0,\mu _{crit}(T),T)$ and $n_{M}(T)=n(M(T),\mu
_{crit}(T),T)$
 coexist as determined by Eqations(\ref{eq_M},\ref{eq_nvsM}). The region of
phase coexistence is shown in Fig.1. For comparison with the MF
solutions we performed Monte-Carlo simulations of the model Eq.(3)
in absence of the Coulomb forces. Due to strong fluctuations in 2D
the critical temperature determined from MC simulations is reduced
by factor of $\sim 2$ in comparison to the MF result.

\section{Coulomb frustrated first order phase transition}

Let us now consider the role of the Coulomb repulsion. The area
under the $ T_{crit}(n)$ in Fig.1 is the area of phase
coexistence. If we fix the temperature the two phases with the
bulk concentrations $n_{0}$ and $n_{M}$ will have volume fractions
$1-x$ and $x$ respectively, where $ x=(n-n_{0})/(n_{M}-n_{0})$.
Since the system is globally electroneutral, the phases with
$n_{0}$ and $n_{M}$ are charged. However, to break
electroneutrality requires a large increase of the Coulomb energy.
As a consequence growth of charged regions with two different
charge densities is blocked by the Coulomb forces.

In the literature there are a few examples of introduction of
charging effects in the problem of phase separation
\cite{MerteljKabanov2005,spivak,ortix,muratov}. There are several
different possibilities to include long-range Coulomb forces in
the model. Muratov\cite{muratov} proposed that the order parameter
is a charged scalar and  the charge density is proportional to the
order parameter. This situation is similar to the problem of a
charged Bose gas in magnetic field considered in \cite{kabal}.
Similar situation is considered in Ref.\cite{ortix} where the free
energy has two distinct minima as a function of the density and
gradient terms in the free energy are replaced by the surface
tension. Jamei, Kivelson and Spivak \cite{spivak} considered the
case with a scalar order parameter where the charge density is
coupled to the order parameter as an external field.

In our case, symmetry allows coupling of the charge density
with square of the order parameter only. Let us consider the
classical free energy density corresponding to the first order phase
transition:
\begin{equation}
F_{1}=((t-1)+(\eta ^{2}-1)^{2})\eta ^{2}\text{.}  \label{fe1}
\end{equation}
Here $t=(T-T_{c})/(T_{0}-T_{c})$ is the dimensionless temperature.
At $t=4/3$ ($ T=T_{0}+(T_{0}-T_{c})/3$) a nontrivial minimum in
the free energy appears. At $t=1$ ($T=T_{0}$) the first order
phase transition occurs. Below $t=1$ the trivial solution $\eta
=0$ corresponds to the metastable phase. At $t=0$ ($ T=T_{c}$) the
trivial solution becomes unstable. In order to study the case of
the Coulomb frustrated phase transition we have to add coupling of
the order parameter to the local charge density. In our case the
order parameter describes the sublattice orbital magnetization and
therefore only square of the order parameter can be coupled to the
local charge density $\rho $:
\begin{equation}
F_{coupl}=-\alpha \eta ^{2}\rho \text{.} \label{fe2}
\end{equation}
The proposed free energy functional is similar to that proposed in
the Ref. \cite{spivak}. In our case the charge plays a role of the
local temperature, while in Ref.\cite{spivak} there is a linear
coupling of the charge to the order parameter and the charge
density plays a role of the external field.

\smallskip The total free energy density should also contain the gradient
term and the electrostatic energy:
\begin{eqnarray}
F_{grad}+F_{el}&=&C(\nabla \eta )^{2}+\frac{1}{2}K[\rho
(\mathbf{r})-\bar{\rho}]\nonumber\\ & &\int d \mathbf{r}^{^{\prime
}}[\rho (\mathbf{r}^{^{\prime }})-\bar{\rho}]/|\mathbf{
r-r}^{^{\prime }}|\text{.}  \label{fe3}
\end{eqnarray}
Here we write $\bar{\rho}$ explicitly to take into account global
electroneutrality. The total free energy
Eqs.(\ref{fe1},\ref{fe2},\ref{fe3}) should be minimized at fixed
$t$ and $\bar{\rho}$.

Next, we proceed to show that the Coulomb term leads to phase
separation in 2D. Minimization of $F$ with respect to the charge
density $\rho ( \mathbf{r})$ leads to the following equation:
\begin{equation}
-\alpha \nabla _{3D}^{2}\eta ^{2}=4\pi K [\rho
(\mathbf{r})-\bar{\rho}]d\delta (z)\text{,}
\end{equation}
here we write explicitly that electrostatic field is 3D but the
charge density $\rho (\mathbf{r})$ is confined in the 2D plane
($z=0$) and depends only on 2D vector $\mathbf{r}$. To preserve
correct dimensionality we introduce the layer thickness d. We
believe that this condition is favorable for creating charge
segregation because electrostatic field is not screened in the
third direction. Solving this equation by applying the Fourrier
transform and substituting the solution back into the free energy
density we obtain:
\begin{eqnarray}
F&=&F_{1}-\alpha \eta ^{2}\bar{\rho}+C(\nabla \eta )^{2}-
\nonumber\\ & &\frac{\alpha ^{2}}{ 8\pi ^{2}K d}\int
d\mathbf{r}^{^{\prime }}\frac{\nabla (\eta (\mathbf{r}
)^{2})\nabla (\eta (\mathbf{r}^{^{\prime
}})^{2})}{|\mathbf{r-r}^{^{\prime }}|}\text{.}
\end{eqnarray}
As a result the free energy functional is similar to the case of
first order phase transition with a shifted critical temperature
due to presence of the term $\alpha \eta ^{2}\bar{\rho}$ and with
additional nonlocal gradient term.

To demonstrate that the uniform solution has a higher free energy
then a nonhomogeneous solution we make the Fourier transformation
of the gradient term:
\begin{equation}
F_{grad}\propto Ck^{2}|\eta _{\mathbf{k}}|^{2}-\frac{\alpha
^{2}k|(\eta ^{2})_{\mathbf{k}}|^{2}}{4\pi K d}\text{,}
\end{equation}
where $\eta _{\mathbf{k}}$ and $(\eta ^{2})_{\mathbf{k}}$ are
Fourier components of the order parameter and square of the order
parameter respectively. If we assume that the solution is uniform
i.e. $\eta _{0}\neq 0 $ and $(\eta ^{2})_{0}\neq 0$ small
nonuniform corrections to the solution reduce the free energy at
small $\mathbf{k}$, where the second term dominates.

The situation is different in 3D. Direct solution of the
equation for the charge density leads to the $local$ gradient term of
higher order $-\frac{\alpha^2}{8\pi K}(\nabla
\eta(\mathbf{r})^{2})^{2}$. This term  can also lead to instability
and higher order expansion in gradient terms become important.

\section{Monte Carlo Simulations}

To substantiate above arguments we performed Monte-Carlo (MC)
simulations of the system described by the Hamiltonian Eq.(3) with
and without the presence of the long-range CR. The simulations
were performed on a square lattice with dimensions $L\times L$
sites with $10\leq L\leq 100$ at different dimensionless
temperatures $t=$ $k_{B}T\epsilon _{0}a/e^{2}$. The short range
potential $v_{l}(\mathbf{i})=V_{l}(\mathbf{i})\epsilon
_{0}a/e^{2}$ \thinspace was taken to be nonzero only for $\left|
\mathbf{i}\right| <2$ and was therefore specified by two
parameters: $v_{l}(1,0)$ and $v_{l}(1,1)$ . To further minimize
the number of free parameters only the nearest neighbors
attractive interaction potential $v_{l}(1,0)$ was taken to be
nonzero in most cases presented here.

We first performed MC simulations of the model at a constant
chemical potential in the absence of CR. Due to presence of first
order phase transition, the particle density probability
distribution $P_{t,\mu }(n)$ has two peaks when the chemical
potential is near the critical value $\mu _{crit}(t)$. At $\mu
_{crit}(t)$ the two peaks have equal height corresponding to the
densities of the two coexisting phases, $n_{0}$ and $n_{M}$. A
standard Metropolis algorithm\cite{Metropolis53} in combination
with simulated annealing\cite{Kirkpatrick83} and histogram
reweighting technique\cite {FerrenbergSwendson1988} gave reliable
results only at higher temperatures near maximum $t_{crit}(n)$. To
improve reliability at lower temperatures we used a variant of
multicanonical aproach\cite{BergNeuhaus1991} adapted to uniformly
sample states over the full range of densities\cite
{OrkoulasPanagiotopulos1998}, $n$, at a constant dimensionless
temperature $ t_{sim}$ and chemical potential $\mu _{sim}$. At
each temperature the final histogram aquisition run involved at
least 10$^{6}$ MC pseudospin flips per site. The density
probability histograms $P_{t_{sim},\mu }(n)$ for several values of
the chemical potential $\mu $ close to the simulation chemical
potential $\mu _{sim}\,$were then calculated at each $t_{sim}$ by
reweighting \cite{FerrenbergSwendson1988} (see Fig. 2). From the
histogram with equal peak heights the densities of the two
coexisting phases $n_{0}$ and $n_{M}$ were then determined at
given $t_{sim}$.

In simulations at constant $n$ one MC step consisted from a single
update per each particle, where the trial move consisted from
setting $S_{z}=0\,$ at the site with nonzero $Q_{\mathbf{i}}$ and
$S_{z}=\pm 1$ at a randomly selected site with zero
$Q_{\mathbf{i}}$. A standard Metropolis algorithm
\cite{Metropolis53} in combination with simulated
annealing\cite{Kirkpatrick83} was used in this case. A typical
simulated annealing run consisted from a sequence of MC
simulations at different temperatures. At each temperature
equilibration phase consisting from $10^{3}-10^{6}$ MC steps was
first executed followed by the averaging phase consisting from the
same or greater number of MC steps. Observables were measured
after each averaging MC step during the averaging phase only.

At constant $n$ in absence of the CR global phase separation below
$ t_{crit}(n)$ occurs in the form of a large cluster with $M\neq
0$. To detect onset of clustering we measure the nearest neighbor
density correlation function (CF) $g_{\rho
L}=\frac{1}{4n(1-n)L^{2}}\sum_{\left| \mathbf{m}\right|
=1}\left\langle \sum_{\mathbf{i}}\left(
Q_{\mathbf{i}+\mathbf{m}}-n\right) \left( Q_{\mathbf{i}}-n\right)
\right\rangle _{L}$, where $\left\langle {}\right\rangle _{L}$
represents the MC average. We define the temperature $ t_{cl}(n)$
at which $g_{\rho L}$ rises to 50\% of its low temperature value
(see Fig. 1b) as the characteristic crossover temperature related
to the formation of clusters.

In Fig. 1a we show the results of MC simulations in absence of the
Coulomb repulsion. We find that for $n\gtrsim 0.4$ the boundary
conditions strongly affect the $t_{crit}(n)$ line calculated at
the constant chemical potential. When we use open boundary
conditions (OBC) $t_{crit}(n)$ is strongly supressed above
$n\approx 0.4$ in comparison to the result obtained with the
periodic boundary conditions (PBC). At constant $n$, on the other
hand, the influence of the boundary conditions on $t_{cl}(n)$ is
less pronounced. The $t_{cl}(n)\,$calculated with both types of
boundary conditions closely follow the $t_{crit}(n)$ line
calculated with PBC. Above $n\approx 0.6$ $ t_{cl}(n)\,$for OBC is
only slightly higer than for PBC. We attribute insensitivity of
$t_{cl}(n)$ to boundary conditions at fixed $n$ to sensitivity of
the correlation function to the short range correlations which are
less sensitive to boundary conditions.
\begin{figure}
\begin{center}
\includegraphics[angle=-0,width=0.45\textwidth]{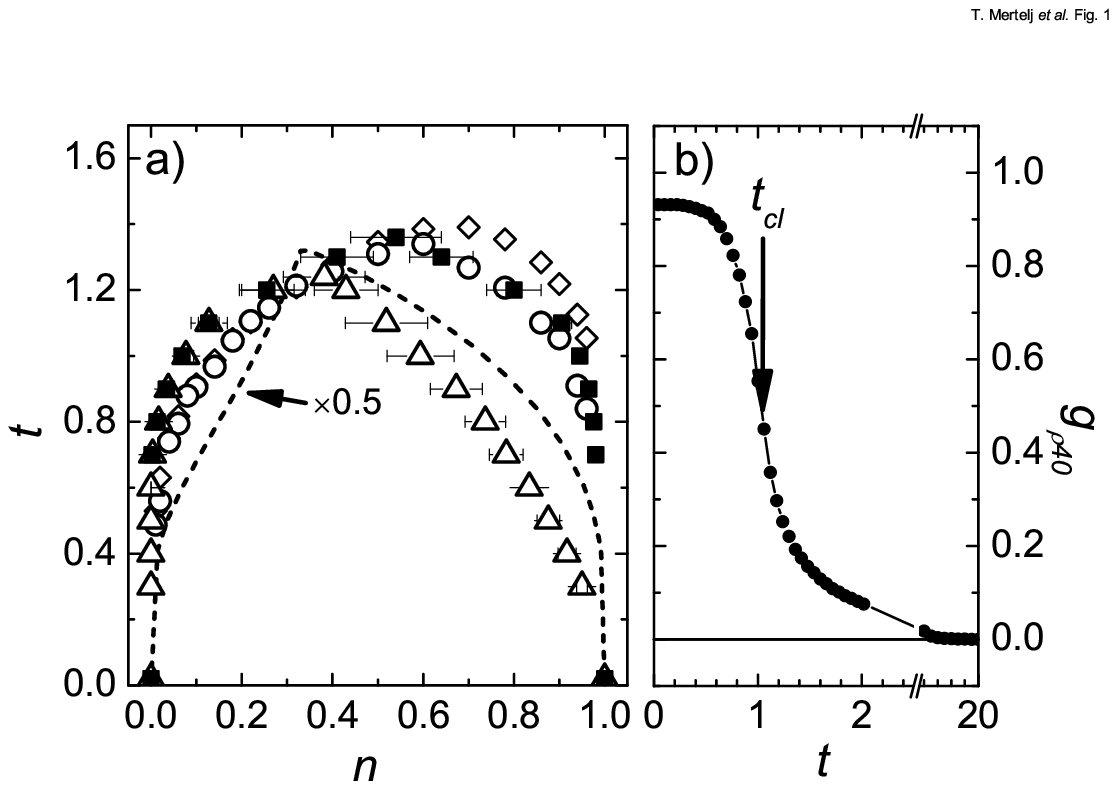}
\vskip -0.5mm \caption{(a) The phase diagram of the model in
absence of the Coulomb repulsion. The dashed line represents the
mean field (MF) solution. The full sqaures ($\blacksquare $)
represent the $t_{crit}(n)$ line calculated for the periodic
boundary conditions by means of the multicanonical MC algorithm
with the system of size $L=40.$ For comparison the $t_{crit}(n)$
line is shown ($\triangle $) for open boundary conditions. The
$t_{cl}(n)$ lines for periodic ($\square $) and open (\large
$\Diamond $\normalsize) boundary conditions are also shown. (b)
The dependence of the nearest neighbor density correlation
function, $g_{\rho L}$, on temperature in absence of the Coulomb
repulsion. The definition of $t_{cl}(n)$ is indicated by arrow.
The numerical errorbars are of the order of symbol sizes}
\end{center}
\end{figure}

Next we analyze the model in the presence of the long-range CR at
constant $n$. In Fig. 3 we show a typical temperature dependence
of the average energy per particle estimator $\left\langle
e_{MC}\right\rangle _{L}$ for different system sizes $L$ in
presence of the long range CR using OBC. Error bars represent the
standard deviation $\sigma _{e_{MC}}=\sqrt{\left\langle
e_{MC}^{2}\right\rangle _{L}-\left\langle e_{MC}\right\rangle
_{L}^{2}}$. The average energy monotonously drops with decreasing
temperature. The drop is more pronounced in the temperature
interval $\sim 0.5>t>\sim 0.1$ in which clusters start to form.
Below $t\sim 0.1$ the clusters are partially ordered. The
temperature dependence of $\left\langle e_{MC}\right\rangle
_{L}\,$is virtually identical for all $L$ (We should note that the
curves are vertically shifted by 0.1 for clarity.) indicating that
the boundary effects on $\left\langle e_{MC}\right\rangle _{L}$
are negligible even for the smalest system sizes.
\begin{figure}
\begin{center}
\includegraphics[angle=-0,width=0.45\textwidth]{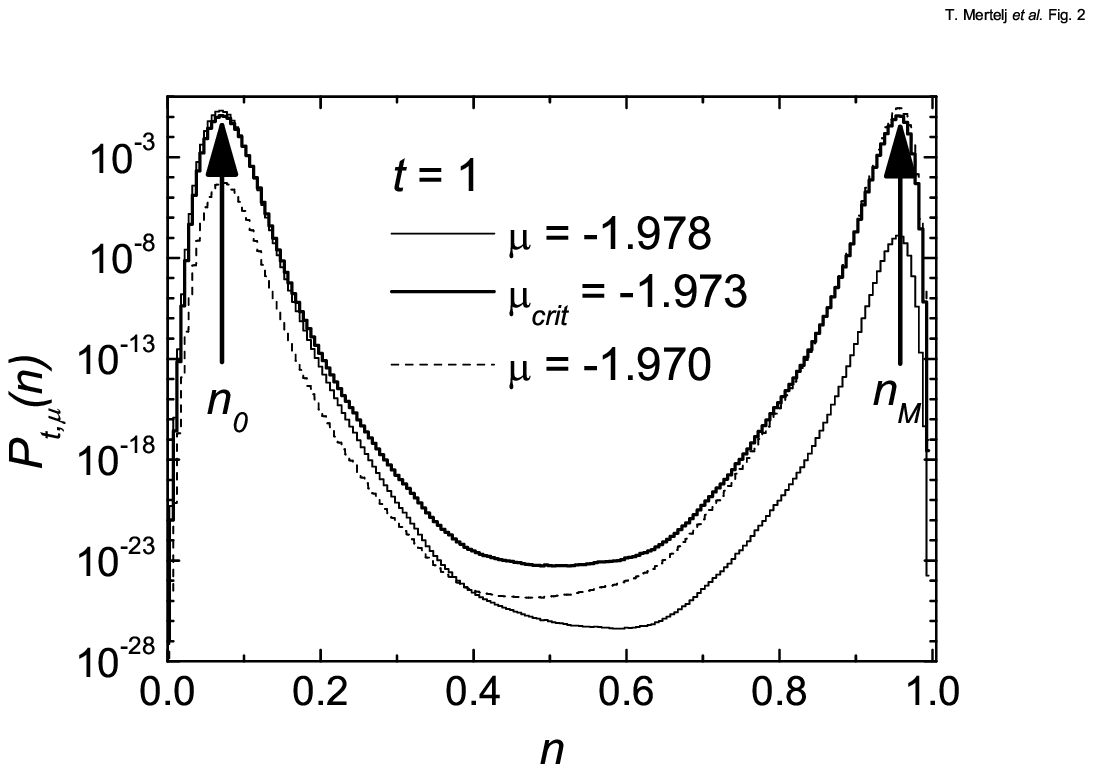}
\vskip -0.5mm \caption{Histograms of the density probability
distribution $P_{t,\mu }(n)$ at the chemical potential near $\mu
_{crit}(t)$ obtained by multicanonical MC simualtion in absence of
Coulomb repulsion The values of the coexisting densities $n_0$ and
$n_M$ at the given temperature are indicated by arrows. Note the
logaritmic scale.}
\end{center}
\end{figure}
\begin{figure}
\begin{center}
\includegraphics[angle=-0,width=0.45\textwidth]{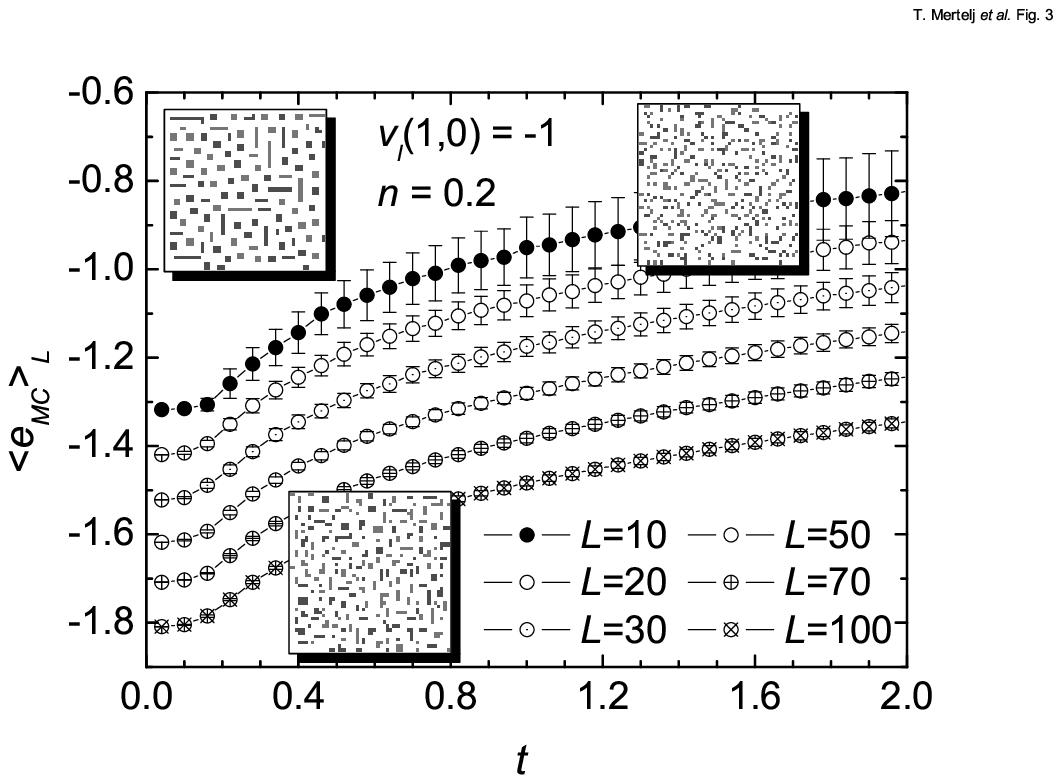}
\vskip -0.5mm \caption{A typical temperature dependence of the
average energy per particle estimator $\left\langle
e_{MC}\right\rangle _{L}$ for different system sizes $L$. Insets
show snapshots of particle distribution at different temperatures,
where darker and brighter shades of grey represent $S_{\mathbf{
i}}^{z}=1$ and $-1$ respectively. Curves are vertically shifted
for 0.1, error bars represent $\sigma _{e_{MC}}$.}
\end{center}
\end{figure}

To check reliability of our simulations we analysed MC update
dynamics by calculating the autocorrelation function of energy
fluctuations,
\begin{eqnarray}
g_{eL}(\tau _{MC})&=&\frac{1}{K\sigma
_{e_{MC}}^{2}}\sum_{i=1}^{K}( e_{MC}( i+\tau _{MC}) \nonumber\\
& &-\left\langle e_{MC}\right\rangle _{L}) ( e_{MC}( i)
-\left\langle e_{MC}\right\rangle _{L}),
\end{eqnarray}
where $e_{MC}\left( i\right) \,$represents the energy per site at
i-th MC step and $\tau _{MC}\,$represents the MC time. A typical
time dependence of $ g_{eL}(\tau _{MC})$ is shown in the inset in
Fig. 4. The autocorrelation function drops with the characteristic
MC relaxation time $\tau _{R}$. $ 1/\tau _{R}$ displays Arhenius
temperature dependence (see Fig. 4) down to the temperature where
clusters start to order. Below this temperature $\tau _{R}$
behaves more erraticaly. The activation energy strongly depends on
the magnitude of the short range potential $v_{l}(1,0)$.
\begin{figure}
\begin{center}
\includegraphics[angle=-0,width=0.45\textwidth]{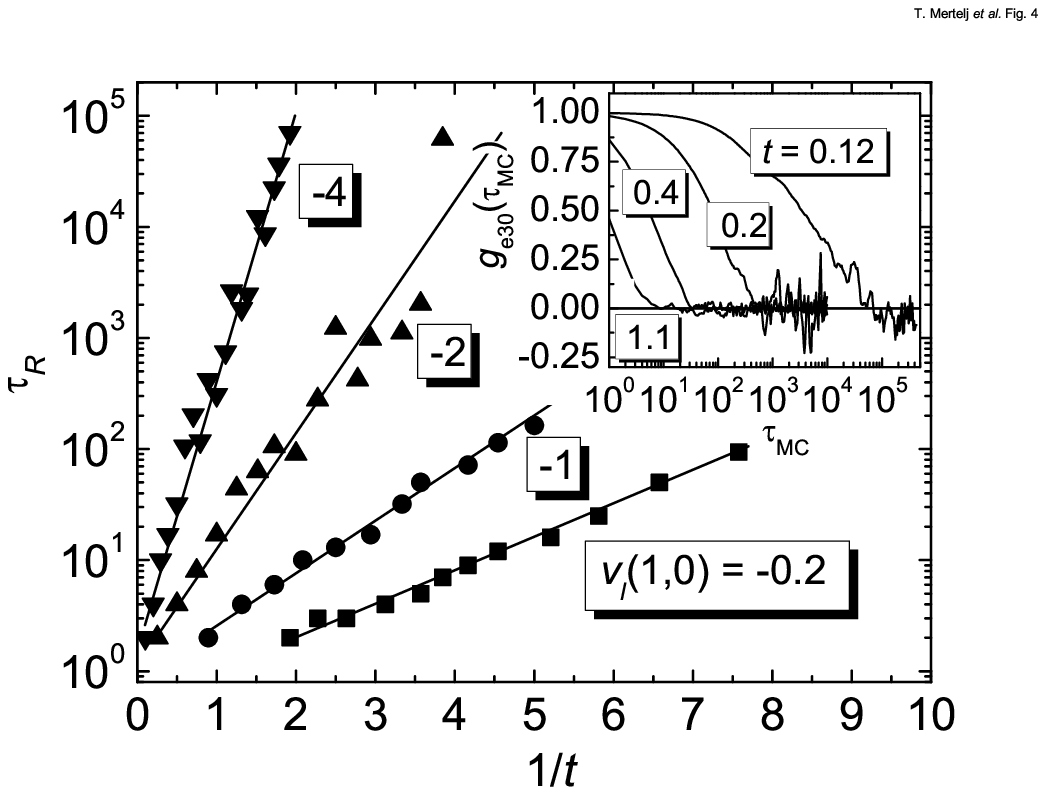}
\vskip -0.5mm \caption{The characteristic MC relaxation time $\tau
_{R}$ as a function of temperature for different values of
$v_{l}(1,0)$. Thin lines represent the Arrhenius fits. The inset
shows the autocorrelation function $g_{e30}(\tau _{MC})$ at a few
temperatures for $v_{l}(1,0)=-1$. For convenience $\tau _{R} $ is
defined as a value of $\tau _{MC}$ where $g_{eL}(\tau
_{MC})=0.25$.}
\end{center}
\end{figure}

In the temperature region where clusters partially order the heat
capacity $ c_{L}=\partial \left\langle e_{MC}\right\rangle
_{L}/\partial t$ displays the peak at $t_{co}(n)$ (see Fig. 5b).
The peak displays no scaling with $L$ indicating that no long
range ordering of clusters appears. Inspection of the particle
distribution snapshots at low temperatures\cite
{MerteljKabanov2005} reveals that finite size domains form (see
Fig. 6). Within the domains the clusters are ordered. The domain
wall dynamics seems to be much slower than our MC simulation
timescale preventing domains to grow. The effective $L$ is
therefore limited by the domain size. This explains the absence of
the scaling and clear evidence for a phase transition near $
t_{co}(n)$. From the simulations is therefore not clear whether
the absence of complete cluster ordering is due to the finitenes
of the MC simulation or it is also due to the glassy form of the
free energy landscape. The square shape of the sample my frustrate
the cluster orders with nontetragonal symmetries, while
practically achievable number of MC steps per temperature step
warrant reliable MC averages only above the temperature which is
of the same order as $t_{co}(n)$. The cluster ordering temperature
$t_{co}(n)\,$
which is the lowest energy scale at all densities is only weakly $n$%
-dependent between $0.1\lesssim n\lesssim 0.9$.
\begin{figure}
\begin{center}
\includegraphics[angle=-0,width=0.45\textwidth]{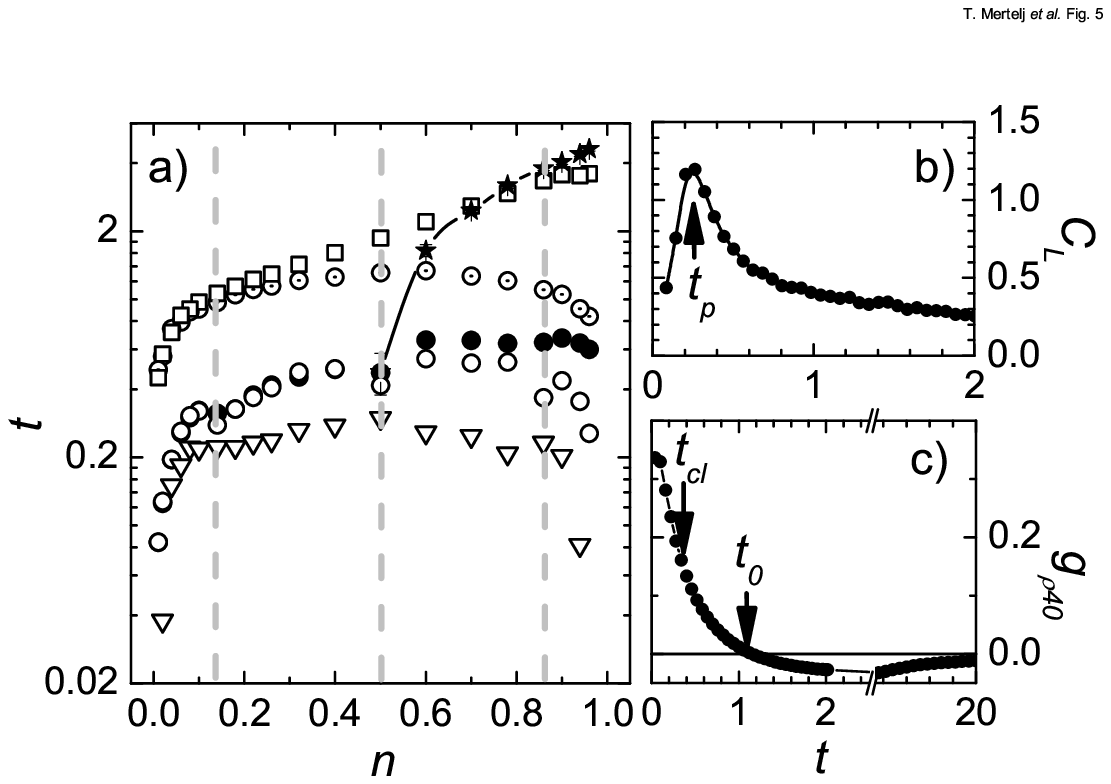}
\vskip -0.5mm \caption{(a) The phase diagram of the model in
presence of the long range Coulomb repulsion. Open circles
({$\circ $}) and full circles ({$\bullet $}) represent the
$t_{cl}(n)$ line for periodic and open boundary conditions
respectively while dotted circles ( $\odot $) represent the
$t_{cl}(n)$ line for periodic boundary conditions in absence of
the long range CR. The onset of clustering, $t_{0}(n)$ is shown by
open squares ($\square $) and the cluster ordering temperature
$t_{CO}(n)$ by open triangles ($\triangle $). The pseudospin
(orbital) ordering temperature is shown by full stars ($\bigstar
$). Note the logarithmic scale. The error-bars are of the order of
symbol sizes or smaller.}
\end{center}
\end{figure}

We now focus on the short range potential shape which promotes
formation of stripes\cite{MerteljKabanov2005}. We set
$v_{l}(1,0)=-1$ and $v_{l}(1,1)=0$ and study the dependence of
clustering on particle density. To detect clustering we again use
the nearest neighbour CF. In Fig. 5c we plot a typical nearest
neighbor CF, $g_{\rho 40}(1,0)$, as a function of temperature. At
high temperatures $t>>$ $\left| v_{l}(1,0)\right| $ CF is slightly
negative due to the long range CR. When the temperature decreases
CF becomes positive and further rises with the decreasing
temperature. No saturation of CF as in the case of absence of the
CR forces is observed with the decreasing temperature (see Fig.
1b). Again we define the temperature at which CF rises to 50\% of
its low temperature value as the characteristic crossover
temperature, $t_{cl}(n)$, related to the formation of clusters.
The dependence of $t_{cl}(n)$ on the particle density is shown in
Fig. 5a for different boundary conditions. While in absence of the
long-range CR $t_{cl}(n)$ closley follows the $t_{crit}(n)$ line
(Fig 1a), suppression of clustering by the CR forces results in a
significant decrease of $t_{cl}(n)$.
\begin{figure}
\begin{center}
\includegraphics[angle=-0,width=0.45\textwidth]{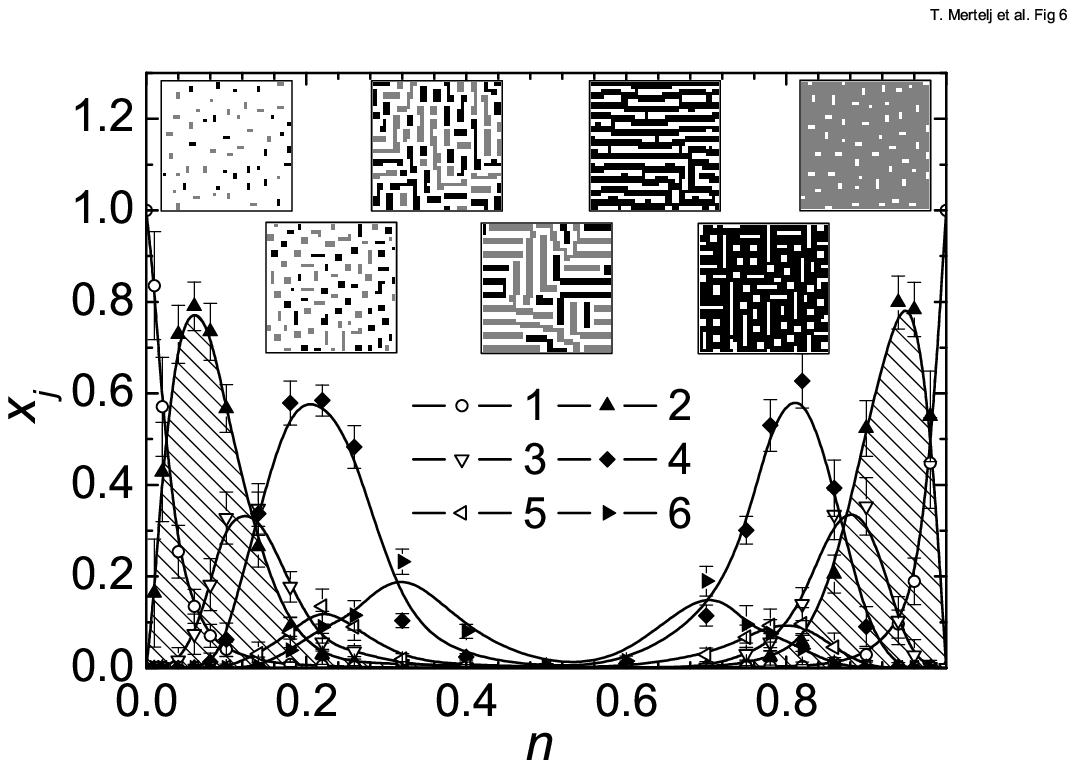}
\vskip -0.5mm \caption{The density dependence of the cluster-size
distribution function $x_{j}$ for a few smallest cluster sizes as
a function of the average density at the temperature $t=0.14$. The
regions of densities where pairs prevail are shadowed. }
\end{center}
\end{figure}

Different boundary conditions influence $t_{cl}(n)$ only for
densities above $n\gtrsim 0.5$. In this region the particles that
form clusters are holes ($Q_{\mathbf{i}}=0$) in the background of
pseudospins ($Q_{\mathbf{i}}=1$). The open boundary conditions are
effectively a perimeter formed form holes which attracts holes and
by pinning enhances hole clustering resulting in an increase of
$t_{cl}(n)$ for OBC.

In addition for $n\geq 0.5$ and our choice of $v_{l}(\mathbf{i})$
the pseudospin background ferromagnetically orders at $t_{S}(n)$
which increases with increasing density as shown in Fig. 5a. The
pseudospin ordering temperature is significantly higher than $
t_{cl}(n)$. Despite this the particle-hole symmetry of the
$t_{cl}(n)$ line is absent. The absence of the the particle-hole
symmetry is a consequence of different entropy contributions of
doubly degenerate particle level ($S_{ \mathbf{i}z}=\pm 1$ for
$Q_{\mathbf{i}}=1$) and singly degenerate hole level
($S_{\mathbf{i}z}=0$ for $Q_{\mathbf{i}}=0$).

The $t_{cl}(n)$ line does not appear smooth. There are clear dips
at $ n\approx 0.14$, $n=0.5$ and $n\approx 0.86$. With increasing
density the ground state of the system apparently goes through a
series of crossovers related to the most probable cluster sizes as
shown in Fig. 6. While the dip at half filling clearly corresponds
to commensurate ordering of stripes the other two dips
approximately correspond to the densities at which clusters of
size four start to replace pairs (see Fig. 6.). There is no
obvious commensuration to underlying lattice at these densities.
At densities at which larger clusters start to replace fours no
comparable anomaly is observed in the $t_{cl}(n)$ line.

Despite the presence of the CR forces some clusters already start
to form at temperatures higher than $t_{cl}(n)$. We can estimate
the upper limit for the onset of cluster formation by the
temperature, $t_{0}(n)$, at which $
g_{\rho L}(1,0)$ crosses 0. It is interesting that $t_{0}(n)$%
\thinspace almost coincides with the $t_{crit}(n)$ line (see Fig.
5a) below $ n\lesssim 0.4$ while at higher densities the onset of
clustering appears at much higher temperatures. In the region
$0.5<n\lesssim 0.75$ the onset of clustering is higher in
temperature than the pseudospin ordering temperature $t_{S}(n)$
while above $n\approx 0.75$ the pseudospin ordering represents the
highest energy scale.

To get further insight in the cluster formation we measured the
cluster-size distribution function. In Fig. 6 we show the low
temperature density
dependency of the cluster-size distribution function, $%
x_{j}=N_{p}(j)/(nL^{2})$, where $N_{p}(j)$ is the number of
particles for $ n\leq 0.5$ or holes for $n>0.5$ in clusters of
size $j$. At the highest temperature $x_{j}$ is close to the
distribution expected for the random ordering. When the
temperature decreases the number of larger clusters starts to
increase at the expense of the single particle number.\cite
{MerteljKabanov2005} Further down in temperature depending on the
average density $n$ clusters of a certain size start to prevail at
the expense of all other sizes. Depending on the particle density
prevailing clusters can be pairs up to $n\approx 0.14$, quadruples
up to $n\,\approx 0.3$ etc.. The situation is qualitatively
symmetrical for the clusters formed by holes at $ n>0.5$. We
should note that for the given $v_{l}(1,0)$ the system prefers
clusters with an even number of particles, however different
$v_{l}(i,j)\,$ might lead to the preference for an odd number of
particles in a cluster.

It should also be emphasized that the preference to certain
cluster sizes becomes clearly apparent only at temperatures lower
then $t_{cl}(n)$, however the transition is not abrupt but gradual
with decreasing temperature. This is seen also from gradual
increase of the average cluster size with decreasing temperature
shown in Fig. 7. Around half filling the average cluster size
starts to diverge at low temperatures indicating formation of long
stripe-like objects (see insets in Fig. 6) and proximity of the
percolation.
\begin{figure}
\begin{center}
\includegraphics[angle=-0,width=0.45\textwidth]{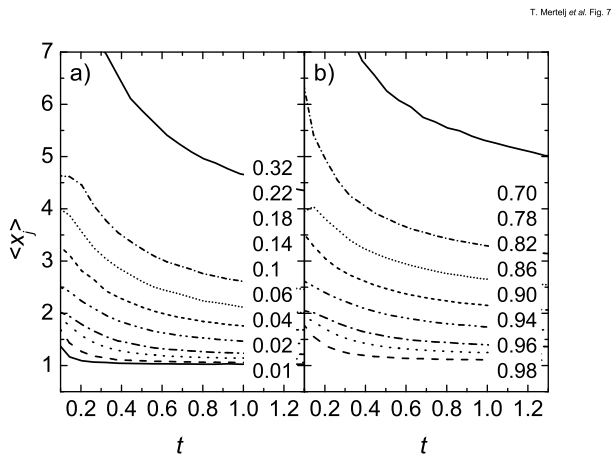}
\vskip -0.5mm \caption{TThe temperature dependence of the average
cluster size for different particle densities $n$ below half
filling (particle clusters) (a) and above half filling  (hole
clusters) (b).}
\end{center}
\end{figure}

\section{Conclusions}

We presented the results of extensive investigation of the
ordering of charged Jahn-Teller polarons as a function of doping
and temperature. We consider charged particles on a 2D square
lattice subject to \textit{only} the long-range Coulomb
interaction and an anisotropic Jahn-Teller (JT) deformation.

We prove that without the long-range Coulomb repulsion the system
is unstable with respect to the first order phase transition below
the density dependent critical temperature. This was demonstrated
by the solution of the mean field equation as well as by direct
Monte Carlo simulations. It was shown that this result does not
depend on the type of boundary conditions and the error due to
finite size effect is estimated.

In the presence of the Coulomb repulsion the global phase
separation becomes unfavorable and the system shows a mesoscopic
phase separation, where the size of the charged regions is
determined by the competition between the ordering energy and the
Coulomb energy. The phenomenological theory of this effect was
formulated where the square of the order parameter is coupled with
the charge density. The charge density plays the role of the local
temperature. This type of coupling is more general in comparison
with the models where the charge plays the role of an external
field.

Using Monte-Carlo (MC) simulations we showed that below a
characteristic clustering temperature the system forms many
different mesoscopic textures, such as clusters and stripes,
depending only on the magnitude of the Coulomb repulsion compared
to the anisotropic lattice attraction and the density of charged
particles. Below the clustering temperature the system goes
through a series of crossovers between phases with different
mesoscopic textures when the particle density is increased. The
low temperature part of the phase diagram is rather symmetric with
respect to half filling. However, above half doping another high
temperature scale appears corresponding to orbital ordering of the
particles. Surprisingly, a feature arising from the anisotropy
introduced by the Jahn-Teller interaction is that objects with an
even number of particles more stable than those with an odd number
of particles. Such a behaviour could have significant implications
for superconductivity when tunnelling is included\cite{mkm}.

\end{document}